\documentclass[a4paper]{jpconf}
\usepackage{graphicx}
\begin{document}
\title{Thermal dileptons as QCD matter probes at SIS}

\author{F Seck$^1$, T Galatyuk$^{1,2}$, R Rapp$^3$ and J Stroth$^{2,4}$}

\address{$^1$ IKP, Technische Universit\"at Darmstadt, Schlossgartenstr. 9, 64298 Darmstadt, Germany}
\address{$^2$ GSI Helmholtzzentrum f\"ur Schwerionenforschung, Planckstr. 1, 64291 Darmstadt, Germany}
\address{$^3$ Cyclotron Institute, Texas A\&M University, College Station (TX) 77843-3366, USA}
\address{$^4$ IKF, Goethe-Universit\"at, Max-von-Laue-Str. 1, 60438 Frankfurt, Germany}

\ead{f.seck@gsi.de}

\begin{abstract}
Electromagnetic radiation is emitted during the whole course of a heavy-ion collision and can escape from the collision zone without further 
interactions. This makes it an ideal tool to study the properties of hot and dense QCD matter. To model the space-time evolution of the collision at 
SIS energies a coarse-graining approach is used to convert transport simulations into meaningful temperatures and densities. These parameters serve 
as input for the determination of the pertinent radiation of thermal dileptons based on an in-medium $\rho$ spectral function that describes 
available spectra at ultrarelativistic collision energies. The resulting excitation function of the thermal excess radiation provides a baseline for 
future measurements by the HADES and CBM experiments at GSI/FAIR, and experiments proposed at NICA and J-PARC.
\end{abstract}

\section{Introduction}

Heavy-ion collisions (HICs) provide insights into the states of matter under extreme densities and temperatures. While the conditions in 
the fireball produced in ultrarelativistic collisions resemble the early universe a few microseconds after the big bang, the matter created in 
collisions at lower bombarding energies might be similar to the conditions in the center of two merging neutron stars \cite{Kurkela_2016, 
Hanauske_2017}. Thus, changing the energy and the type of the nuclei which induce the reactions permits systematic investigations of the properties 
and the composition of matter across the QCD phase diagram.

For example, chiral symmetry which is spontaneously broken in the QCD vacuum due to a non-zero quark-antiquark condensate $\langle\bar{q}q\rangle$ 
gets restored in a hot medium \cite{Borsanyi_2010, Bazavov_2012}.
This induces modifications to the spectral distributions of chiral partners in the light hadron sector such as the $\rho$ and $a_1$ mesons -- 
ultimately leading to their degeneracy.
A substantial melting of the chiral condensate is conjectured to be relevant already for HICs at low collision energies \cite{Klimt_1990, 
Schaefer_2007}. 

Electromagnetic (EM) radiation represents an excellent probe for such investigations as it is emitted during the whole time evolution of a HIC 
and decouples from the strongly interacting collision zone once created. In contrast to real photons which are massless, virtual photons decaying 
into a pair of leptons ($e^+e^-$ or $\mu^+\mu^-$) have an additional observable in terms of their invariant-mass. Thus, one can access the 
information about the properties of matter produced inside the hot and dense fireball which is irretrievable from the spectra of final-state hadrons 
due to rescattering.

In particular, the excess yield of low-mass dileptons above the contribution from the hadronic freeze-out cocktail was identified to be sensitive 
to the fireball lifetime, while the slope in the intermediate-mass region of the dilepton invariant-mass spectrum can serve as a thermometer which is
not distorted by the collective expansion of the medium~\cite{Rapp_vanHees_2016, Specht_2010}.

\section{Coarse-grained transport approach for thermal dilepton rates}

For a proper theoretical description of the contribution of in-medium signals to the dilepton invariant-mass spectrum realistic thermal dilepton 
emission rates need to be convoluted with an accurate modeling of the chemical potentials and temperatures reached during the space-time evolution of 
the fireball.

For HICs at ultrarelativistic energies (URHICs) a hydrodynamic description can be used to evaluate the thermodynamic properties of the medium and the 
thermal dilepton rates can directly be employed. The applicability of such an approach for the system evolution down to the SIS18 energy 
range of a few GeV is unclear. One issue is the justification of thermalization due to the long time it takes until a full overlap of the 
two incoming nuclei is reached.
For this reason hadronic transport models like UrQMD are commonly used to describe the system evolution of HICs at SIS energies. The dilepton 
radiation is then obtained perturbatively by integrating for each created resonance the probability of a decay into a dilepton over the whole 
lifetime of this hadronic resonance. The incorporation of medium effects on broad resonances into these off-equilibrium approaches remains however 
challenging.

\begin{figure}[ht]
\vspace{0.3cm}
\begin{minipage}[t]{0.47\textwidth}
\hspace{0.1cm}%
\includegraphics[width=0.97\textwidth]{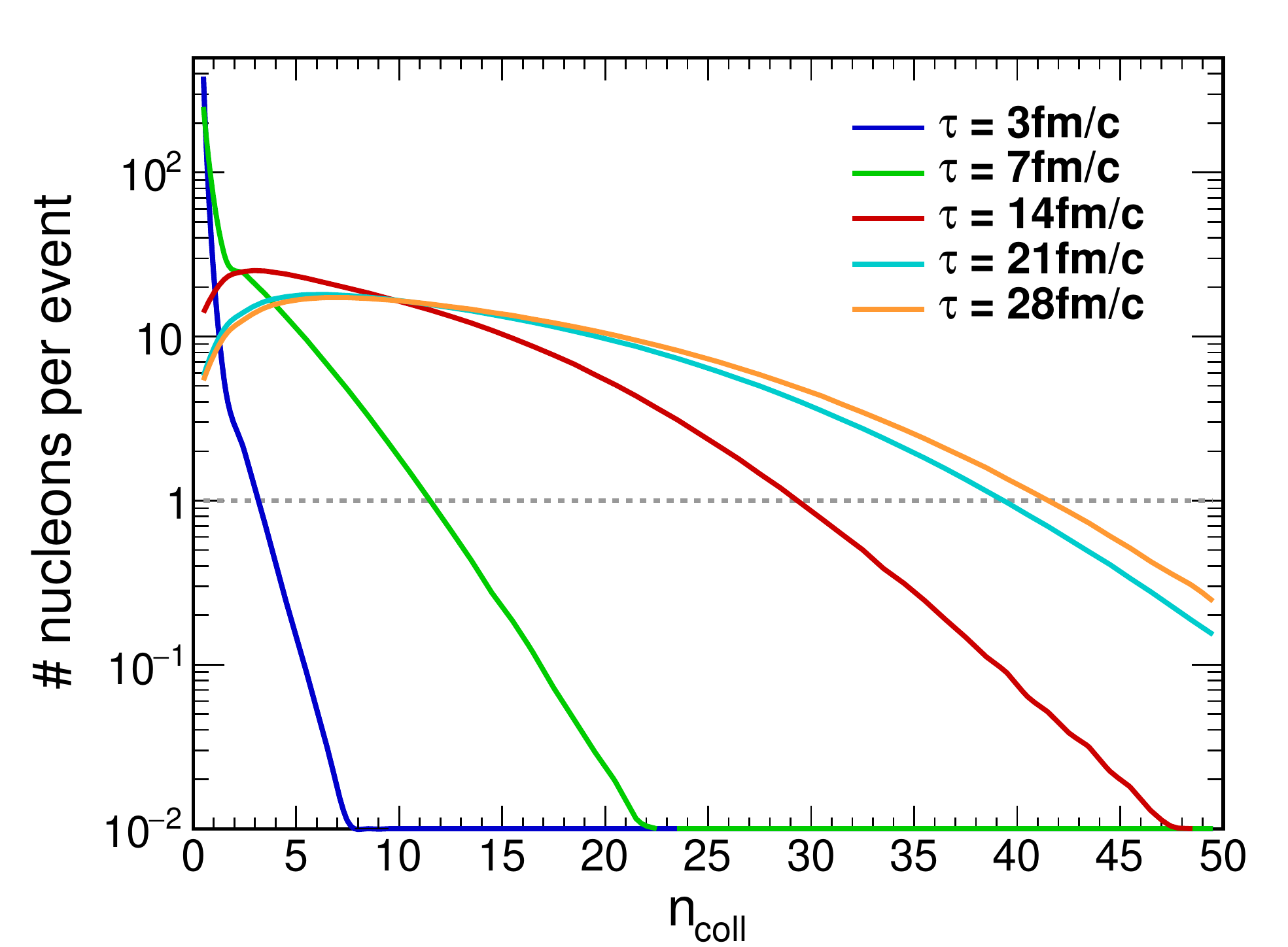}
\caption{Distribution of the number of binary collisions of nucleons per event in central Au+Au collisions at 
$\sqrt{s_{NN}}$=2.4~GeV for different time steps in the evolution.}
\label{fig:ncoll}
\end{minipage}\hspace{0.8cm}%
\begin{minipage}[t]{0.47\textwidth}
\hspace{0.15cm}%
\includegraphics[width=0.95\textwidth]{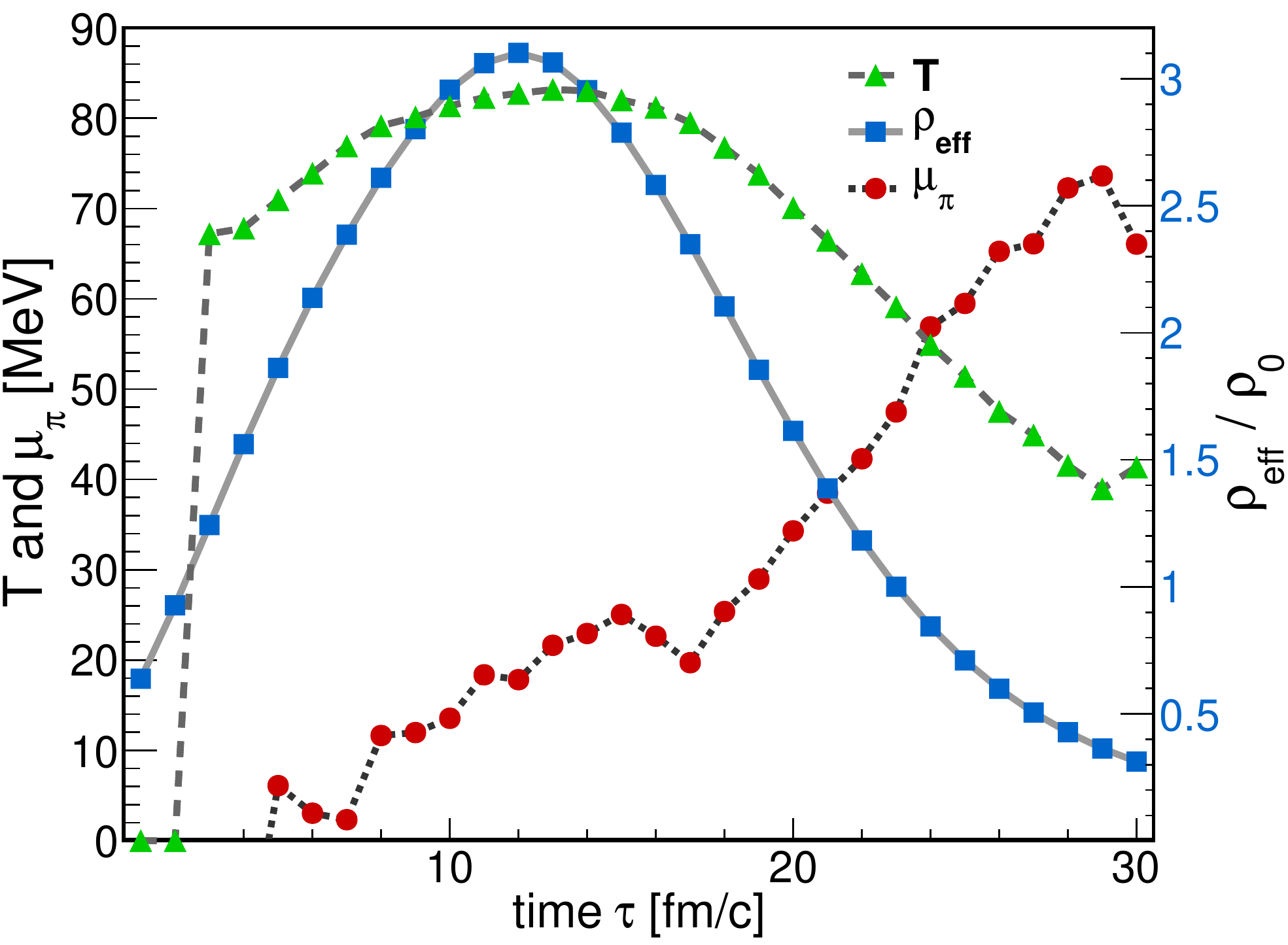}
\caption{Time evolution of temperature, effective baryon density (right scale) and effective pion chemical potential averaged 
over an inner cube of 7x7x7 cells (1\,fm$^3$ each) in central Au+Au collisions at $\sqrt{s_{NN}}$=2.4~GeV.}
\label{fig:temp_dens}
\end{minipage} 
\end{figure}

To bridge this gap between the microscopic transport and macroscopic hydrodynamic approaches a coarse-graining procedure was 
proposed~\cite{Huovinen_2002}. It was recently applied to the SIS18 energy regime~\cite{Endres_2015, Galatyuk_2016, Seck_2017} where the HADES 
collaboration has measured the dielectron spectra in collisions of Ar+KCl at $E_{\rm lab}$=1.76A\,GeV and Au+Au at $E_{\rm lab}$=1.23A\,GeV 
(corresponding to $\sqrt{s_{NN}}$=2.6 and 2.4~GeV respectively)~\cite{HADES_2011, Kornakov_2017}. By dividing the space-time evolution into 
4-dimensional cells and averaging over an ensemble of many simulated transport events one obtains smooth particle distributions. Reasonable 
temperatures, baryon and pion densities as well as collective flow patterns can then be extracted from cells which fulfill certain criteria that are 
favorable for (the onset of) thermalization, \textit{i.e.}, a minimum number of collisions that the nucleons in the cell need to have undergone so 
that their momentum distributions in all Cartesian directions acquire a Gaussian shape with comparable width as well as the transverse mass spectra 
of pions that can be described by an exponential shape. In this way the premise of a full hydro simulation is mitigated as the deviations from the 
vanishing mean-free path limit are kept in the evolution. 
Figure~\ref{fig:ncoll} shows how the distribution of nucleons in central Au+Au collisions at $\sqrt{s_{NN}}$=2.4~GeV which have experienced a given 
number of collisions is modified over time. During the first few time steps the distribution changes quite rapidly due to many interactions in the 
system, while it stays almost the same after 21~fm/c indicating kinetic freeze-out. The evolution of the extracted temperature, effective baryon 
density 
$\rho_{\textit{eff}} = \rho_{N}+\rho_{\bar{N}} + \frac{1}{2}(\rho_{R}+\rho_{\bar{R}})$ ($N$ refers to nucleons, $R$ to baryonic resonances) for 
central collisions averaged over the inner cube of 7x7x7 cells (each of a volume of 1\,fm$^3$) is shown in Fig.~\ref{fig:temp_dens}. In the center 
of the Au+Au collision system temperatures of up to 90 MeV and densities of up to 3 times normal nuclear matter density are reached.

These bulk properties of the cells are used as input parameters for the calculation of thermal dilepton radiation using the expression  
\begin{equation}\label{eq:emissivity}
\frac{d^8N}{d^4x\,d^4p} = \frac{\alpha^2_{EM}}{\pi^2\,M^2}\,f_B(p_0,T)\,\varrho_{EM}(M,p;T,\rho_{\textit{eff}},\mu_{\pi}) \, , 
\end{equation}
where $M=\sqrt{p_0^2-p^2}$ is the invariant mass of the virtual photon, $f_B$ denotes the thermal Bose distribution function, and $\varrho_{EM}$ the 
EM spectral function of the QCD medium depending on the temperature, the effective baryon density, and an effective chemical potential of pions, 
$\mu_{\pi}$. For $\varrho_{EM}$ we employ a parametrization of the in-medium $\rho$ spectral function of Ref.~\cite{Rapp_Wambach_1999} which 
describes dilepton data in URHICs.
The effect of the chemical potential of pions on the EM spectral function is encoded in an overall fugacity factor, 
$z^{\kappa}=\exp(\mu_{\pi}/T)^{\kappa}$, where $\kappa$ specifies the average number of pions participating in the production of $\rho$ mesons. Using 
UrQMD we estimated this number to be close to 1 at SIS energies~\cite{Galatyuk_2016} as the main production channels are baryon resonance decays 
rather than $\pi\pi$ annihilation.

\section{Results}

\begin{figure}[ht]
\begin{minipage}[t]{0.47\textwidth}
\hspace{0.3cm}%
\includegraphics[width=0.9\textwidth]{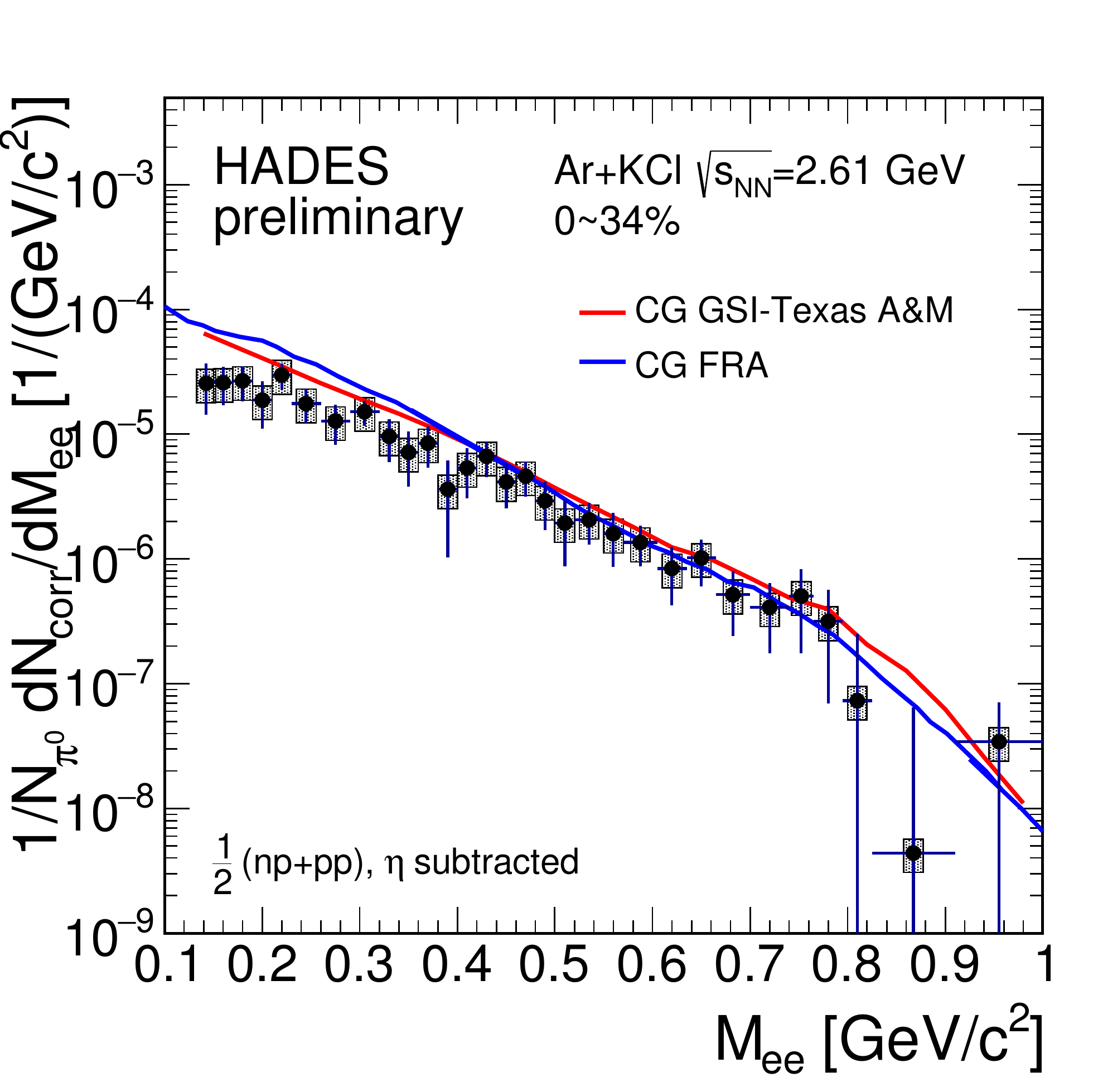}
\caption{Comparison of dilepton excess spectra from two independent coarse-graining approaches~\cite{Endres_2015,Galatyuk_2016} and the 
experimentally extracted yield above the freeze-out cocktail in Ar+KCl at 1.76AGeV from HADES~\cite{HADES_2011,HADES_2010}.}
\label{fig:arkcl}
\end{minipage}\hspace{0.8cm}%
\begin{minipage}[t]{0.47\textwidth}
\hspace{0.3cm}%
\includegraphics[width=0.88\textwidth]{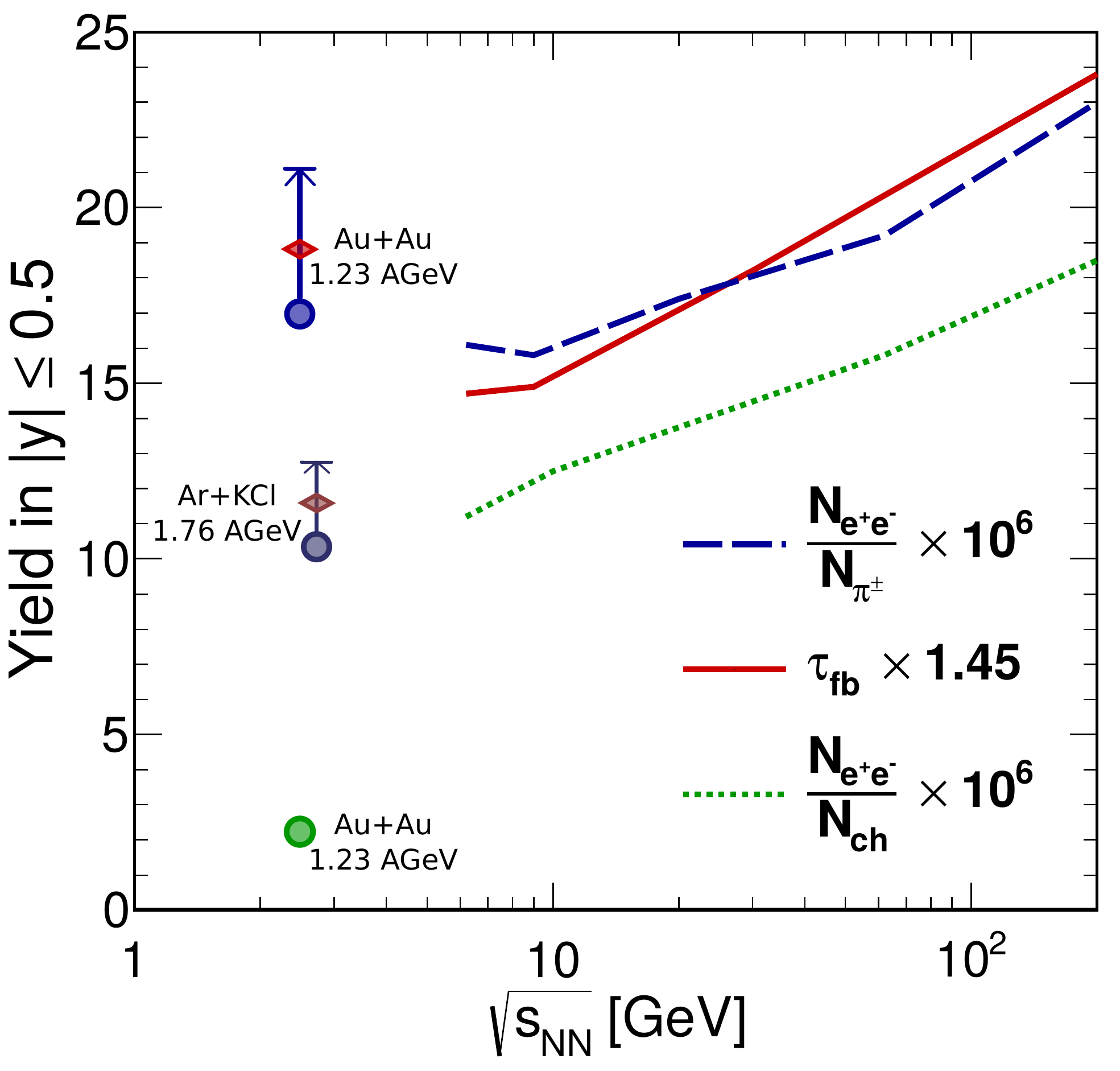}
\caption{Excitation function of the thermal low-mass dilepton yield in central Au+Au collisions normalized to $N_{\pi^\pm}$ (blue dashed line, blue 
circle) and $N_{\rm ch}$ (green dotted line, green circle), together with the lifetime of the system (red solid line, red diamond).}
\label{fig:excite_func}
\end{minipage} 
\end{figure}

An experimental acceptance filter was applied to the full phase-space distribution in order to compare the resulting invariant-mass spectrum of 
thermal radiation to available data of the Ar+KCl system. Figure~\ref{fig:arkcl} shows a fair agreement between the calculated spectrum inside the 
HADES acceptance and the measured excess yield above the ``cocktail'' of long-lived EM decays at freeze-out plus a superposition of p+p and n+p 
reactions~\cite{HADES_2011,HADES_2010} to model the contribution of first-chance NN collisions.
The high baryon densities reached during the collisions lead to strong medium modifications of the spectral shape of the $\rho$ meson resulting in an 
almost exponential excess spectrum of thermal radiation.

One remarkable finding is that the cumulative yield of thermal dileptons in the low-mass region, $M_{ee}$=0.3-0.7\,GeV/c$^{2}$, reveals the same 
temporal development as the build-up of the collective radial flow~\cite{Galatyuk_2016, Seck_2017}. This underlines a close relation of these 
quantities as the same interactions which drive the collectivity in the system also excite resonances which in turn emanate dileptons. It therefore 
consolidates the use of the low-mass dilepton yield as a fireball chronometer. The interacting fireball lifetime amounts to approximately 13\,fm/c in 
central Au+Au collisions at $\sqrt{s_{NN}}$=2.4~GeV and to about 8\,fm/c for the lighter Ar+KCl system.

The quantitative link between the fireball lifetime and the thermal dilepton yield established for URHICs~\cite{Rapp_vanHees_2016} normalizes the 
latter to the number of charged particles, $N_{\rm ch}$, at midrapidity. To generalize this finding also to lower collision energies one has to keep 
in mind that $N_{\rm ch}$ is no longer a good proxy for the thermal excitation energy in the system as the fireball is more and more dominated by the 
incoming nuclei. Thus, Fig.~\ref{fig:excite_func} shows the low-mass thermal dilepton yield normalized to the number of charged pions, $N_{\pi^\pm}$. 
The curves have been calculated using a fireball model with QGP and in-medium hadronic emission~\cite{Rapp_vanHees_2016}, while the diamonds and 
circles represent the yields and lifetime obtained within our coarse-graining approach. The proportionality of the yield to the lifetime for URHICs 
stays intact with a slightly larger normalization of 1.45 (which includes strong final-state decays in the dilepton yield)~\cite{Galatyuk_2016}

\begin{equation}\label{eq:yield_to_lifetime}
\left. \frac{N_{ll}}{N_{\pi^\pm}}\right|_{|y| \leq 0.5} \times 10^6 = 1.45 \, \tau_{\rm fb} \, [{\rm fm}/c].
\end{equation}

The excitation function remains rather flat from top RHIC down to the SIS18 energies. This offers the opportunity to utilize the measurement of the 
thermal low-mass yield in the search for the phase transition in the so-far unexplored (with dileptons) FAIR energy regime. A critical slowing down 
of the evolution in the transition region could manifest itself by an increased lifetime of the fireball and in turn an enhanced dilepton yield.

\begin{figure}[ht]
\includegraphics[width=0.47\textwidth]{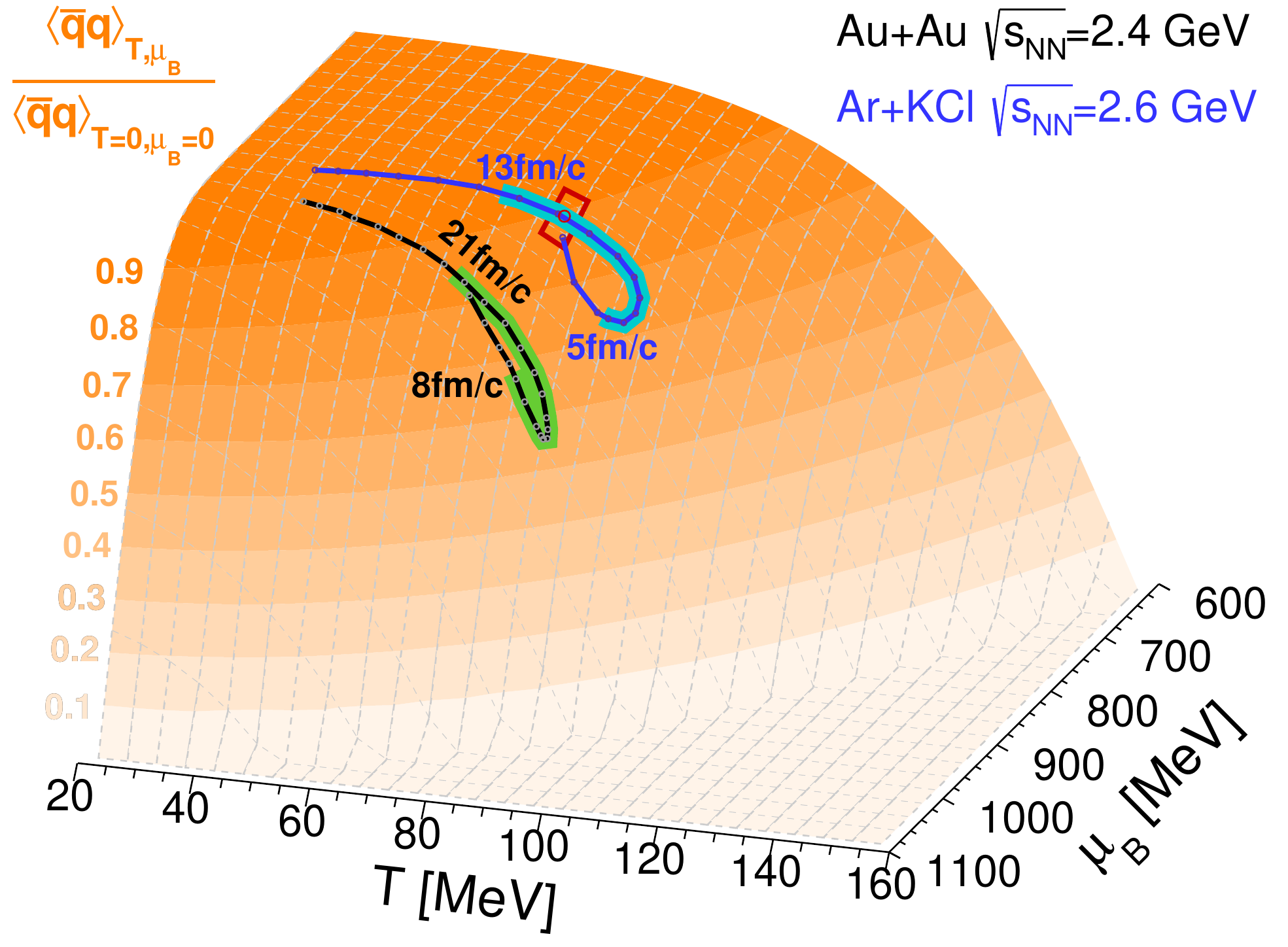}\hspace{0.3cm}%
\begin{minipage}[b]{0.51\textwidth}
\caption{Trajectories of the central regions of Au+Au and Ar+KCl collisions in the QCD phase diagram. The orange contour lines indicate the reduction 
of the chiral condensate compared to the vacuum expectation value based on a NJL model calculation~\cite{Schaefer_2007}. The green (cyan) highlighted 
part of the trajectory of Au+Au (Ar+KCl) shows the time period which radiates most of the thermal dilepton yield. The chemical freeze-out point of 
the 
Ar+KCl system~\cite{HADES_2011b} is depicted by the red square. \\ $\left. \right.$}
\label{fig:qcd_diag}
\end{minipage}
\end{figure}

Translating the extracted temperatures and baryon densities to the baryon chemical potential $\mu_{B}$ one can plot the trajectories in the QCD phase 
diagram of the central regions of Au+Au and Ar+KCl collisions at SIS energies together with the expectation value of the chiral condensate employing 
the Nambu-Jona-Lasinio (NJL) model~\cite{Schaefer_2007}, see Fig.~\ref{fig:qcd_diag}.
The trajectories are traversed in counter-clockwise direction in time steps of 1\,fm/c. The highlighted regions show the time interval from which 
most (90\%) of the thermal dileptons emanate. The chiral condensate during the emission window is depleted significantly. Note that the leading 
density approximation using the sigma term of nucleons suggest a 25-30\% drop already at normal nuclear matter density which is not reflected in the 
model calculation. Shown is also the location of the chemical freeze-out point extracted from a statistical hadronization model (SHM) fit to the 
abundances of several hadron species in Ar+KCl collisions~\cite{HADES_2011b}. Shortly after this point on the trajectory the emission of thermal 
dileptons creases indicating thermal freeze-out.

\section{Conclusions and outlook}

The study of HICs at relativistic bombarding energies becomes increasingly important as results from several theoretical frameworks locate the QCD 
critical point in regions accessible to FAIR~\cite{Fischer_2014, Critelli_2017}. Thermal dilepton emission in HICs at such energies can be obtained 
with the presented coarse-graining approach that couples the space-time evolution of hadronic transport simulations with in-medium dilepton rates.
The procedure was applied to Au+Au and Ar+KCl collision at $\sqrt{s_{NN}}$=2.4~GeV or 2.6~GeV, respectively, and shows a fair agreement with the 
measured dilepton excess. Interesting insights emerge from the close correlation between the build-up of collectivity and the time window of thermal 
dilepton emission which also corroborate the possibility to track the lifetime of the system with the radiation yield in the mass window 
0.3\,GeV/c$^2\,\leq\,M_{ll}\leq$~0.7\,GeV/c$^2$.
This opens the opportunity to apply this approach to the FAIR energy regime where up to now no dilepton measurements exist and to study the impact 
of different scenarios onto the invariant-mass spectrum of thermal lepton pairs.

\ack
We thank the HADES collaboration for providing the data points of the dilepton excess spectrum in Ar+KCl collisions and H.~van Hees and S.~Endres for 
their coarse-graining curve. 
This work was supported by the U.S. National Science Foundation under grant PHY-1614484, the A.-v.-Humboldt Foundation (Germany), the Helmholtz-YIG 
grant VH-NG-823 at GSI and TU Darmstadt (Germany), and the Hessian Initiative for Excellence (LOEWE) through the Helmholtz International Center for 
FAIR (HIC for FAIR).

\section*{References}


\end{document}